# Air passivation of chalcogen vacancies in two-dimensional semiconductors

Yuanyue Liu,* Pauls Stradins, and Su-Huai Wei*

Drs. Y. Liu, P. Stradins, S. Wei[+]
National Renewable Energy Laboratory, Golden, CO, USA
[+]present address: Beijing Computational Science Research Center, Beijing, China
*yuanyue.liu.microman@gmail.com, suhuai.wei@nrel.gov

**ABSTRACT:** Defects play important roles in semiconductors (SCs). Unlike those in bulk SCs, defects in two-dimensional (2D) SCs are exposed to the surrounding environment, which can potentially modify their properties/functions. Air is a common environment; yet its impact on the defects in 2D SCs still remains elusive. In this work, we unravel the interaction between air and chalcogen vacancies ($V_X$)-the most typical defects in 2D SCs. We find that, although the interaction is weak for most molecules in air, $O_2$ can be chemisorbed at $V_X$ with a barrier that correlates with the SC cohesive energy and can be overcame even at room temperature for certain SCs. Importantly, the chemisorbed $O_2$ changes the $V_X$ from commonly-believed harmful carrier-traps to electronically benign sites. This unusual behavior originates from the iso-valence between $O_2$ and X when bonded with metal. Based on these findings, we propose a facile approach to improve the performance of 2D SCs by using air to passivate the defects.

**Main text**: Two-dimensional (2D) semiconductors (SCs) are promising candidates for next-generation electronics, optoelectronics and energy conversion/storage devices.[1] They feature an extremely high surface-to-mass ratio due to their thickness of only one or several atomic layers. As a result, the atoms (including the atomic defects) are exposed to the environment that surrounds the SC, and are subjective to the environmental factors.[2] This has been demonstrated in many experiments, which show that the environment can modify the properties and performance of 2D SCs.[3]

Defects are usually more reactive than the perfect sites and therefore ought to have a stronger coupling with the environment. Since defects play important roles in the properties and performance of materials, it is necessary to evaluate the impact of environment on the defects. Air is a common environment that is involved in material processing and device operation. However, its impact still remains elusive. Specifically, it is poorly understood: what is the strength of the interaction (physical or chemical), what is the effect on the electronic properties, and why is this effect?

Chalcogen vacancies ($V_X$; X = S, Se, and Te) are the most typical defects in 2D SCs.[4] Therefore, in this work, we investigate the interaction between air and $V_X$. We find that, different from other molecules, $O_2$ can be chemisorbed at $V_X$. This chemisorption occurs with a barrier, which correlates with the cohesive energy of the SC and can be overcame even at room temperature for certain SCs. Im-portantly, the chemisorbed $O_2$ changes the $V_X$ from commonly-believed harmful carrier-traps to electronically benign sites. The electronic origin is explained and an effective way to improve the performance of 2D SCs is proposed.

We focus on two types of the most popular 2D SCs—transition metal dichalcogenides ($TMX_2$; TM = Mo and W), and group-III monochalcogenides (IIIX, where III = Ga and In)—which, in the perfect form, have good air stability at room temperature. Both types have a hexagonal unit cell with metal layers located in between two X layers. In contrast to $TMX_2$, IIIX have two metal layers that form III-III bonds. First-principles calculations are performed by using the Vienna Ab-initio Simulation Package (VASP)[5] with projector augmented wave (PAW) pseudopotentials.[6] Unless specified, the Perdew-Burke-Ernzerhof (PBE) exchange-correlation functional[7] is employed. The Heyd-Scuseria-Ernzerhof (HSE) functional (which gives a more accurate approximation of the exchange-correlation energy, but is more computationally expensive)[8] and van der Waals corrections (DFT-D3)[9] are also used in some cases to validate the robustness of the conclusions. A supercell consisting of 6x6 primitive cells is used, and the vacuum along the surface normal direction is kept to be > 15 Å. The plane-wave cut-off energy is 400 eV. All the structures are relaxed until the final force on each atom is < 0.01 eV/Å. The reaction barriers are calculated by using the climbing-image nudged elastic band (CINEB) method.[10]

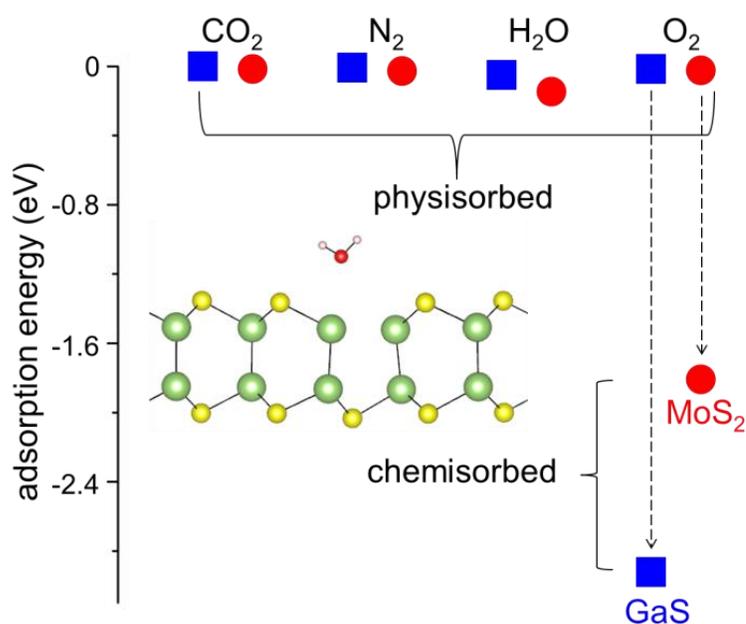

**Figure 1**. Adsorption energies of major molecules in air on $V_X$ for GaS and $MoS_2$. Inset shows the structure of $H_2O$ adsorbed at $V_X$ of GaS. Yellow: S; Green: Ga; Red: O; White: H. $O_2$ has two adsorption states: physisorption and chemisorption.

Figure 1 shows the adsorption energies of the major molecules in air on Vx of GaS and MoS$_2$. The adsorption energy is defined as:

$$E_{ad} = E(V_X+\text{mol}) - E(V_X) - E(\text{mol}), \qquad (1)$$

where $E(V_X+\text{mol})$ is the energy of the system with a molecule adsorbed on the $V_X$ site, and $E(V_X)$ and $E(\text{mol})$ are the energies for isolated systems. Most molecules (including H$_2$O) are only physisorbed, i.e., they are located above the top X layer with a distance > 2 Å (see Fig. 1 inset) and adsorption energies < 0.2 eV. In contrast, O$_2$ can also be chemisorbed at $V_X$ (see structures in Fig. 2), reducing the energy by > 1.5 eV. Adding van der Waals (vdW) corrections to the calculations shows a decrease of physisorption energies, but does not change the preference of adsorption states.

Although physisorbed molecules can modify the electronic structure of the defects by depleting/accumulating the charge densities,[3a, 3b, 3f, 11] this modification is not stable at room temperature because the molecules can easily desorb or diffuse away from the defect sites. In contrast, the chemisorbed O$_2$ cannot be released at ambient conditions due to the strongly exothermic adsorption; therefore, it can irreversibly change the properties of the defects. However, the chemisorption of O$_2$ does not occur spontaneously. As shown in Fig. 2, a barrier must be overcome before reaching the chemisorbed state from the physisorbed state. The calculated barrier heights depend on the computational methods (see Fig. 4); however, it is generally believed that PBE underestimates the barriers and the best estimate is given by the HSE functional.[12] We find from HSE calculations that MoS$_2$ has a larger barrier (0.74 eV) than that of GaS (0.48 eV). The reason why IIIX has a relatively small barrier will be explained later (see Fig. 4 and the related text). Nevertheless, these barriers are insignificant at room temperature. According to the rate theory, the transition time from the physisorbed state to chemisorbed state is:

$$t \sim 1/(f e^{-E_b/k_b T}), \qquad (2)$$

where $f$ is attempt frequency, $E_b$ is the barrier, $k_b$ is the Boltzmann constant, and $T$ is temperature. The $f$ can be estimated by the flux of O$_2$ arriving at the $V_X$:

$$f \sim n \, v \, s_d, \qquad (3)$$

where $n$ is O$_2$ density in air, $v$ is the speed of the O$_2$ molecule, and $s_d$ is the cross section of $V_X$, which can be taken as the square of lattice parameter. At room temperature and one atmospheric pressure, f is ~ $10^8$ molecules/s, which gives $t \sim$ 20 hrs for MoS$_2$ and 2 s for GaS.

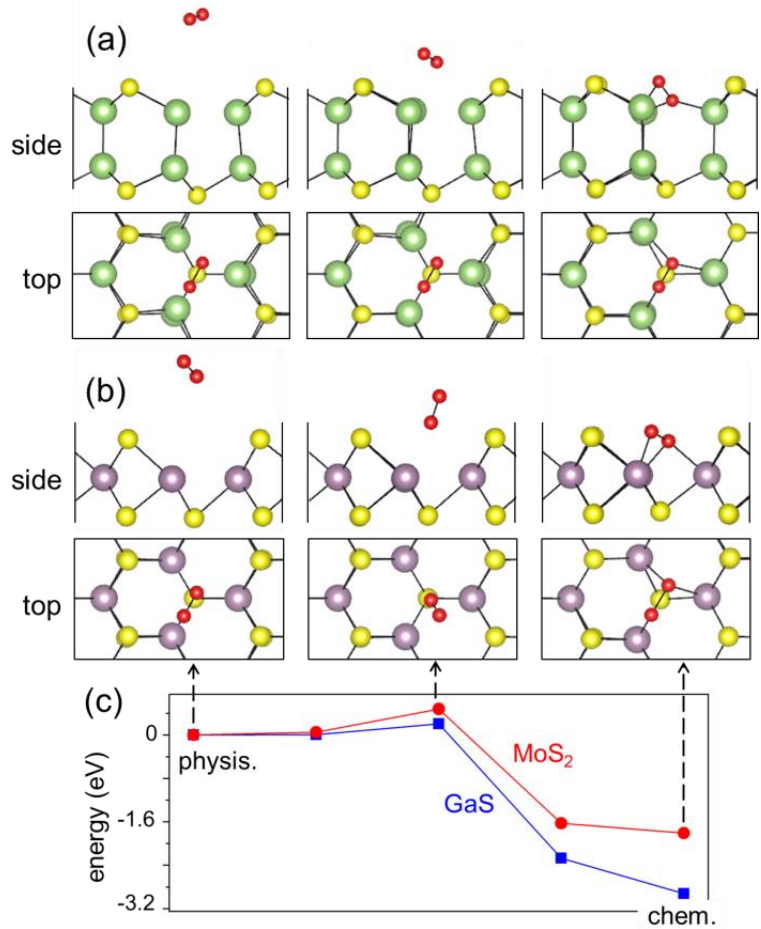

**Figure 2**. Reaction pathway of $O_2$ with $V_X$ in GaS (a) and $MoS_2$ (b), from the physisorbed to chemisorbed state. Yellow: S; Green: Ga; Purple: Mo; Red: O. (c) shows the energy evolution during the reaction, calculated by using PBE. See Fig. 4 for a complete set of calculated barriers.

Chemisorbed $O_2$ has a significant impact on the electronic structure of $V_X$. As shown in Fig. 3, an intact $V_X$ creates both deep acceptor and donor levels in GaS, and a deep acceptor level in $MoS_2$. These levels can trap the electrons or holes and scatter other charge carriers in the SC, thus reducing the carrier mobility. They can also non-radiatively recombine electrons and holes, resulting in a decrease of the quantum efficiency. Hence, $V_X$ are typically harmful to the material's performance (Note that this is different from 2D black phosphorous whose intrinsic defects are relatively electronically benign[13]). However, upon $O_2$ chemisorption, the deep gap states are removed completely (Fig. 3), converting $V_X$ to electronically inactive sites. The elimination of the defect states is due to the isovalence between the $O_2$ and X atom in 2D SCs. $O_2$ molecule has two singly occupied $\pi^*$ orbitals (two unpaired electrons), which can obtain two electrons and make $O_2$ act as an element with the valence of -2, similar to that of the X atom in IIIX and $TMX_2$. Therefore, $O_2$ chemisorption at $V_X$ is equivalent to isovalent substitution of X by $O_2$; thus, the electronic structure of the SCs remains largely intact. As a proof, Fig. 3 inset shows the change of electron density after $O_2$ chemisorption. The electron density difference is defined as:

$$\Delta\rho = \rho(V_X + O_2) - \rho(V_X) - \rho(O_2), \quad (4)$$

where $\rho(V_X + O_2)$ is the electron density of the system with $O_2$ adsorbed on $V_X$ site, $\rho(V_X)$ and $\rho(O_2)$ are the electron densities for isolated systems. We observe that electrons accumulate at $O_2$ by depleting those previously in $V_X$. The electron-accumulating region has a shape similar to that of $\pi^*$, indicating that $\pi^*$ orbitals are now fully occupied and the dangling electrons in $O_2$ are paired up. This is further confirmed by the change of spin from triplet in isolated $O_2$ to singlet after chemisorption. As a result, the system becomes isovalent with that of the perfect SC, and thus, defect states are eliminated. This effect is generally applicable to other TMX2 and IIIX.

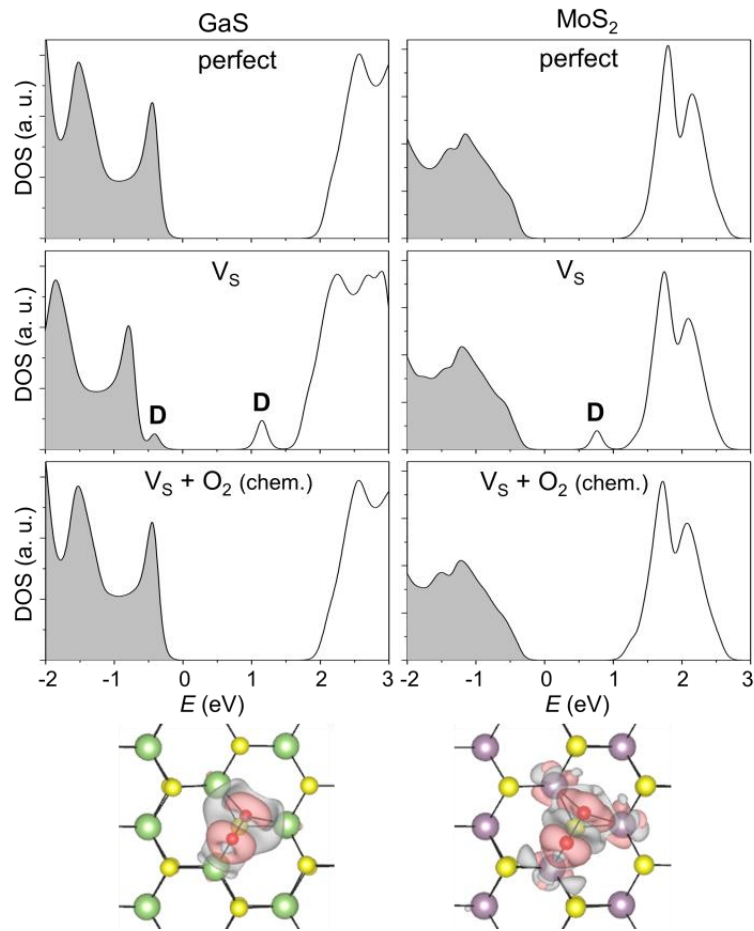

**Figure 3.** Density of states (DOS) for perfect GaS (left) and $MoS_2$ (right), and those containing $V_X$, and $O_2$ chemisorbed at the $V_X$ site. The defect states are marked by "D" labels. At the bottom, we show the isosurface plots of the change of electron density difference after $O_2$ chemisorption, suggesting that electrons are transferred from $V_X$ to the π* orbitals of $O_2$. Gray: electron depletion; Pink: electron accumulation.

Figure 4 shows the trends of the chemisorption barriers across various commonly studied 2D SCs, calculated by using different methods. Note that some of the $TMX_2$ or IIIX are not semiconducting (e.g., $WTe_2$[14]), or do not exist in 2D (e.g., InS), and therefore, they are excluded here. We observe that adding vdW corrections decreases the barriers due to the enhancement of $O_2$ binding with the substrate. HSE always gives the highest values, which are closest to reality.[12] Regardless of the calculation methods, we find that the barriers decrease as the cohesive energy of the SC decreases. This is because the cohesive energy represents the stability of the material—thus, a less "stable" SC has a stronger tendency to chemisorb $O_2$ at $V_X$. For the same cations, the barriers drop from S to Se to Te; and for the same anions, the barriers increase from Ga to Mo to W. Consequently, $WS_2$ has the largest barrier of 0.97 eV, and therefore the $O_2$ chemisoprtion is kinetically forbidden at room temperature.

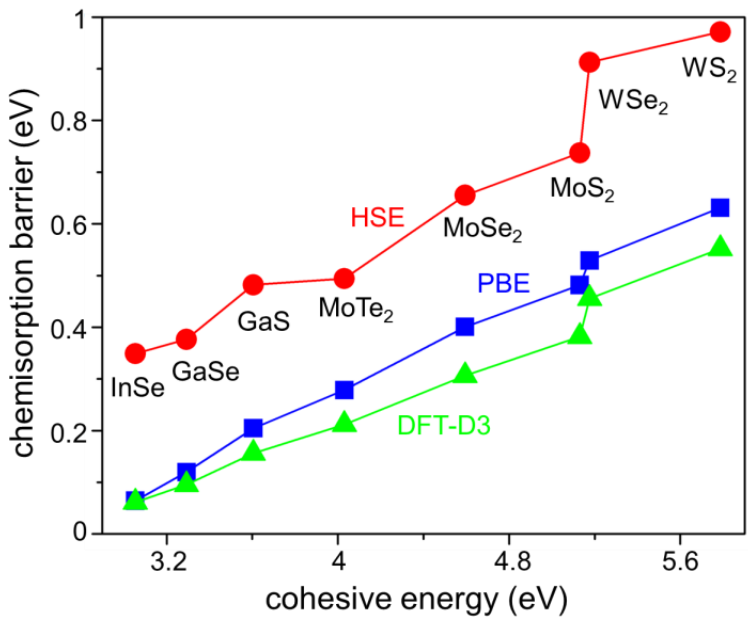

**Figure 4**. Trends of the $O_2$ chemisorption barriers at $V_X$ for various 2D semiconductors, calculated by using different methods (HSE, PBE, and DFT-D3).

Chemisorbed $O_2$ at $V_X$ can further dissociate into separated O atoms: one at the $V_X$ site, and the other bound with the neighboring X to form an O-X group, as shown in the SM Fig. S1. This dissociation also encounters barriers that depend on specific materials (Fig. S1). However, compared with intact $V_X$, there are still no deep gap states (Fig. S2). This is because both O and the O-X group are isovalent with X atoms in these 2D SCs; therefore, the defect states due to the missing X is eliminated, as explained previously for chemisorbed $O_2$.

Our results suggest that it is possible to improve the transport properties and the quantum efficiency of the 2D chalcogenide SCs by $O_2$ treatment. An increase of temperature could help the reaction of $O_2$ with $V_X$ in $WS_2$ and $WSe_2$, which have the highest barriers. However, the physisorbed $O_2$ at other sites of the basal plane can cause the nonuniform variation of the potential, leading to a scattering of the charge carriers [3c, 15]. Therefore, it is important to remove these molecules after treatment; otherwise, they will compensate the improvement of the defect properties.

We note that several recent experimental studies show that the photoluminescence intensity of $TMX_2$ is enhanced after air/O treatment.[3b, 16] One possible mechanism offered in literature[3b, 16] is the charge transfer between the environmental molecules and the $V_X$. Here we propose that these may partially result from the increase of quantum efficiency because of the suppressed non-radiative recombination at the $V_X$, thanks to the elimination of trap states by $O_2$. However, since the physisorbed $O_2$ (or other molecules) at other sites of the basal plane can also modify the optical properties of the $TMX_2$ and contribute to the total photoluminescence variation, making a confirmative conclusion is too early at this stage. Nevertheless, it is encouraging to decouple these two contributions experimentally.

Although here we focus on the 2D chalcogenide SCs, we believe that $O_2$ could also passivate the O vacancies in 2D oxide SCs, or perhaps other vacancies that have the missing atoms isovalent with $O_2$. In fact, the improved performance after air/O treatment has been observed in ZnO thin-film devices.[17]

In summary, by using the chalcogen vacancies as an example, we investigate the interaction between air and the defects in 2D SCs. $O_2$ is found to have the strongest impact that can change the vacancies from harmful carrier-traps to electronically benign sites. The electronic origin of this impact is explained and an effective way to improve the performance of 2D SCs is proposed.

## ACKNOWLEDGMENT

This research was funded by the U.S. Department of Energy (DOE) under Contract No. DE-AC36-08GO28308. Y. L. and P. S. acknowledge support from DOE SETP DE-EE00025783 and DE-EE0006336. This work used the Peregrine supercomputer at National Renewable Energy Laboratory (NREL), as well as the computational resources at National Energy Research Scientific Computing Center (NERSC), which is supported by the Office of Science of the DOE under Contract No. DE-AC02-05CH11231. Y. L. acknowledges computational resources and staff support from Lawrence Livermore National Laboratory (LLNL).

## REFERENCES

[1] aG. Fiori, F. Bonaccorso, G. Iannaccone, T. Palacios, D. Neumaier, A. Seabaugh, S. K. Banerjee, L. Colombo, *Nat. Nanotechnol.* **2014**, *9*, 768-779; bQ. H. Wang, K. Kalantar-Zadeh, A. Kis, J. N. Coleman, M. S. Strano, *Nat. Nanotechnol.* **2012**, *7*, 699-712; cD. Akinwande, N. Petrone, J. Hone, *Nat Commun* **2014**, *5*; dC. N. R. Rao, H. S. S. Ramakrishna Matte, U. Maitra, *Angew. Chem. Int. Ed.* **2013**, *52*, 13162-13185; eY. Hou, A. B. Laursen, J. Zhang, G. Zhang, Y. Zhu, X. Wang, S. Dahl, I. Chorkendorff, *Angew. Chem. Int. Ed.* **2013**, *52*, 3621-3625.

[2] X. Zou, B. I. Yakobson, *Acc. Chem. Res.* **2015**, *48*, 73-80.

[3] aS. Tongay, J. Suh, C. Ataca, W. Fan, A. Luce, J. S. Kang, J. Liu, C. Ko, R. Raghunathanan, J. Zhou, F. Ogletree, J. Li, J. C. Grossman, J. Wu, *Sci. Rep.* **2013**, *3*; bS. Tongay, J. Zhou, C. Ataca, J. Liu, J. S. Kang, T. S. Matthews, L. You, J. Li, J. C. Grossman, J. Wu, *Nano Lett.* **2013**, *13*, 2831-2836; cH. Qiu, L. Pan, Z. Yao, J. Li, Y. Shi, X. Wang, *Appl. Phys. Lett.* **2012**, *100*, 123104; dH.-Y. Park, M.-H. Lim, J. Jeon, G. Yoo, D.-H. Kang, S. K. Jang, M. H. Jeon, Y. Lee, J. H. Cho, G. Y. Yeom, W.-S. Jung, J. Lee, S. Park, S. Lee, J.-H. Park, *ACS Nano* **2015**, *9*, 2368-2376; eD. J. Late, B. Liu, H. S. S. R. Matte, V. P. Dravid, C. N. R. Rao, *ACS Nano* **2012**, *6*, 5635-5641; fB. Chen, H. Sahin, A. Suslu, L. Ding, M. I. Bertoni, F. M. Peeters, S. Tongay, *ACS Nano* **2015**, *9*, 5326-5332; gJ. O. Varghese, P. Agbo, A. M. Sutherland, V. W. Brar, G. R. Rossman, H. B. Gray, J. R. Heath, *Adv. Mater.* **2015**, *27*, 2734-2740; hD. Kiriya, M. Tosun, P. Zhao, J. S. Kang, A. Javey, *J. Am. Chem. Soc.* **2014**, *136*, 7853-7856; iS. S. Chou, Y.-K. Huang, J. Kim, B. Kaehr, B. M. Foley, P. Lu, C. Dykstra, P. E. Hopkins, C. J. Brinker, J. Huang, V. P. Dravid, *J. Am. Chem. Soc.* **2015**, *137*, 1742-1745.

[4] aJ. Hong, Z. Hu, M. Probert, K. Li, D. Lv, X. Yang, L. Gu, N. Mao, Q. Feng, L. Xie, J. Zhang, D. Wu, Z. Zhang, C. Jin, W. Ji, X. Zhang, J. Yuan, Z. Zhang, *Nat Commun* **2015**, *6*; bZ. Yu, Y. Pan, Y. Shen, Z. Wang, Z.-Y. Ong, T. Xu, R. Xin, L. Pan, B. Wang, L. Sun, J. Wang, G. Zhang, Y. W. Zhang, Y. Shi, X. Wang, *Nat Commun* **2014**, *5*.

[5] aG. Kresse, J. Hafner, *Phys. Rev. B* **1993**, *47*, 558-561; bG. Kresse, J. Furthmüller, *Phys. Rev. B* **1996**, *54*, 11169-11186.

[6] aP. E. Blöchl, *Phys. Rev. B* **1994**, *50*, 17953-17979; bG. Kresse, D. Joubert, *Phys. Rev. B* **1999**, *59*, 1758-1775.

[7] J. P. Perdew, K. Burke, M. Ernzerhof, *Phys. Rev. Lett.* **1996**, *77*, 3865-3868.

[8] J. Paier, M. Marsman, K. Hummer, G. Kresse, I. C. Gerber, J. G. Ángyán, *J. Chem. Phys.* **2006**, *124*, 154709.

[9] S. Grimme, J. Antony, S. Ehrlich, H. Krieg, *J. Chem. Phys.* **2010**, *132*, 154104.

[10] G. Henkelman, B. P. Uberuaga, H. Jónsson, *J. Chem. Phys.* **2000**, *113*, 9901-9904.

[11] aS. Zhao, J. Xue, W. Kang, *Chem. Phys. Lett.* **2014**, *595–596*, 35-42; bQ. Yue, Z. Shao, S. Chang, J. Li, *Nanoscale Research Letters* **2013**, *8*, 425.

[12] J. F. Binder, A. Pasquarello, *Phys. Rev. B* **2014**, *89*, 245306.

[13] Y. Liu, F. Xu, Z. Zhang, E. S. Penev, B. I. Yakobson, *Nano Lett.* **2014**, *14*, 6782-6786.

[14] aM. N. Ali, J. Xiong, S. Flynn, J. Tao, Q. D. Gibson, L. M. Schoop, T. Liang, N. Haldolaarachchige, M. Hirschberger, N. P. Ong, R. J. Cava, *Nature* **2014**, *514*, 205-208; bI. Pletikosić, M. N. Ali, A. V. Fedorov, R. J. Cava, T. Valla, *Phys. Rev. Lett.* **2014**, *113*, 216601.

[15] X. Cui, G.-H. Lee, Y. D. Kim, G. Arefe, P. Y. Huang, C.-H. Lee, D. A. Chenet, X. Zhang, L. Wang, F. Ye, F. Pizzocchero, B. S. Jessen, K. Watanabe, T. Taniguchi, D. A. Muller, T. Low, P. Kim, J. Hone, *Nat. Nanotechnol.* **2015**, *10*, 534-540.

[16] H. Nan, Z. Wang, W. Wang, Z. Liang, Y. Lu, Q. Chen, D. He, P. Tan, F. Miao, X. Wang, J. Wang, Z. Ni, *ACS Nano* **2014**, *8*, 5738-5745.

[17] aD. C. Olson, S. E. Shaheen, R. T. Collins, D. S. Ginley, *J. Phys. Chem. C* **2007**, *111*, 16670-16678; bL. Chien Cheng, W. Meng Lun, L. Kuang Chung, H. Shih-Hua, C. Yu Sheng, L. Gong-Ru, H. JianJang, *Display Technology, Journal of* **2009**, *5*, 192-197.

**TOC:**

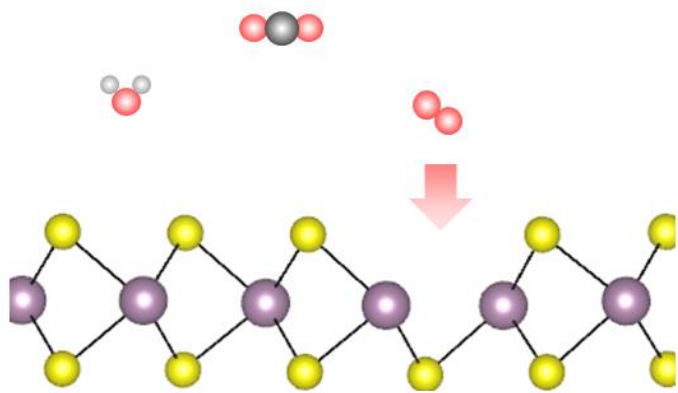